\documentclass[aps,pre,showpacs,amssymb,twocolumn]{revtex4-1}
\usepackage{graphicx,color,psfrag,amsmath,bm}
\newcommand{\alb}[1]{{#1}}

\newcommand{\average}[1]{\left<{#1}\right>}

\newcommand{\pq}[1]{\left[{#1}\right]}

\usepackage{xr}
\begin{document}
%\author{ Hans C. Fogedby\inst{1} \and Alberto Imparato\inst{1}}
%\affiliation
%\institute{\inst{1} Department of Physics and
%Astronomy, University of
%Aarhus, Ny Munkegade\\
%8000 Aarhus C, Denmark\\}
\title{Autonomous quantum rotator}
\author{Hans C. Fogedby and Alberto Imparato}
\affiliation{Department of Physics and
Astronomy, University of
Aarhus, Ny Munkegade\\
8000 Aarhus C, Denmark\\}
\begin{abstract}
%\abstract{
We consider a minimal model of a quantum rotator composed of a single particle confined in an harmonic potential and driven by two temperature-biased heat reservoirs.
In the case the particle potential is rendered asymmetric and rotated an angle, a finite angular momentum develops, corresponding to a directed rotary motion.
At variance with the classical case, the thermal fluctuations in the baths give rise to a non-vanishing average torque contribution;
this is a genuine quantum effect akin to the Casimir effect. In the steady state the heat current flowing between the two baths is systematically converted into particle rotation.
We derive  exact expressions for the work rate and heat currents in the case where the system is driven by an external time periodic mechanical force. We show, in agreement with previous works on classical systems, that for this choice of external manipulation protocol, the rotator 
cannot work either as a heat pump or as  a heat engine.
We finally use our exact results to extend an ab-initio quantum simulation algorithm to the out-of-equilibrium regime.
%}
\end{abstract}

\maketitle
%%%%%%%%%%%%%%%%%%%%%%%%%%%%%%%%%%%%%%%%%%%%%%%%%%%%%%%%%
%%%%%%%%%%%%%%%%%%%%%%%%%%%%%%%%%%%%%%%%%%%%%%%%%%%%%%%%%
%\section{Introduction}
%%%%%%%%%%%%%%%%%%%%%%%%%%%%%%%%%%%%%%%%%%%%%%%%%%%%%%%%%
%%%%%%%%%%%%%%%%%%%%%%%%%%%%%%%%%%%%%%%%%%%%%%%%%%%%%%
There is currently a strong interest in quantum thermal machines \cite{Devoret2014}.
From a theoretical point of view these systems are of interest in elucidating the limits of thermodynamics 
and the role of fluctuations \cite{Kosloff13}.  Experimentally, it has recently been feasible to manufacture nano devices 
acting like  reciprocating heat engines \cite{Rossnagel16} or thermoelectric transducers \cite{Thierschmann2015}. 

%\alb{ {\it Reduce this part to a few lines}
%A heat engine is a device driven by two biased heat reservoirs which is able to act either as an amplifier or
%a refrigerator. The simplest quantum heat engine is a three level amplifier. It was proposed by 
%Scovil and Schulz-DuBois \cite{Scovil1959}. Driving the ground state and the highest level by a hot reservoir 
%(temperature $T_1$) and the ground state and the intermediate level by a cold reservoir (temperature $T_2$)
%in order to maintain population inversion, the model acts like an amplifier when coupled to an external field at 
%resonance. The efficiency of the three level amplifier is $\eta=1-T_2/T_1$ equal to the Carnot efficiency. 
%performance $\epsilon=T_2/(T_1-T_2)$.
%There are no fluctuations in the three level heat engine and thus no dissipation; the levels coupling to the heat 
%baths are assumed to be in equilibrium. Nevertheless, owing to the transitions there is non equilibrium energy 
%flow from the hot reservoir to the cold reservoir. The three level model is  incomplete. It is basically a reversible
%or quasi-static engine and can therefore operate with efficiency only limited by the Carnot efficiency. 
%We note that reversing the operation the three level system acts like a refrigerator with coefficient of 
%performance $\epsilon=T_2/(T_1-T_2)$.}

Small quantum devices are open quantum systems coupled to their environment,
possibly characterised by a time-dependent Hamiltonian. 
%This issue is of great current interest in the context of quantum computing, decoherence, entanglement, etc. 
Consequently, in the modelling of a quantum heat engine or refrigerator one must include the coupling to the heat 
reservoirs subject to a proper characterisation \cite{Breuer02}. %{\it In the classical case assuming a time scale separation the fast degrees of freedom of the heat reservoirs are modelled by white noise and the description  of a small classical system coupled to heat reservoirs is conveniently for example discussed in terms of Langevin equations, alternatively, in terms of Fokker-Planck equations \cite{Zwanzig2001}. In the quantum case one has to  exercise more caution. Strictly speaking the concepts of dissipation and irreversibility are absent in a proper quantum description which requires a Hamiltonian framework. In other words,}
More precisely, and different from the classical case \cite{Zwanzig2001}, the heat baths have to  be characterised explicitly as
larger quantum systems coupled to the small quantum system \cite{Ford65,Ford87,Ford88,Breuer02}.

Many efforts have been devoted to the investigation of quantum reciprocating engines, performing e.g.,  Carnot,  Otto or  Stirling cycles. These cycles require an external agent that changes periodically one or more mechanical parameters in the Hamiltonian and the system temperature, see e.g.,  \cite{Gelbwaser13,Rossnagel16,Insinga16} or  \cite{Kosloff13} and references therein.
The simplest quantum equivalent to the Carnot cycle is a heat engine  proposed by 
Scovil and Schulz-DuBois \cite{Scovil1959}: it is a three level amplifier operating in contact with two heat reservoirs.  This model acts like an amplifier when coupled to an external field at 
resonance \cite{Gesuic1967}. However it is basically a reversible
or quasi-static engine and can therefore operate with efficiency only limited by the Carnot efficiency.
\alb{Another example of thermal motor is the flywheel introduced  in \cite{Levy16} which can store kinetic energy when driven by an external time-periodic field and when monitoring and feedback control are applied.}

Another class of thermal machines are the quantum autonomous  devices which exhibit the ideal design for engineering purposes since they can operate in steady state conditions without any external
time dependent drive. While quantum autonomous engines such as refrigerators have been investigated, see e.g., \cite{Brunner16}, quantum autonomous motors exhibiting directed transport have not attracted much attention.

An autonomous rotor model has recently been proposed in \cite{Roulet17}. In this  model the strength of the coupling with the two heath baths depends on the state of the system itself. \alb{ Another interesting example is the quantum mill \cite{Seah18} whose working fluids are two qbits at different temperatures  that can perform work against a  dissipative load.}
Here we show that an autonomous rotor can be obtained with a much simpler design.

A minimal model of a classical autonomous heat  motor exhibiting directed rotary motion at a non vanishing rate was first proposed by  Filliger and Reimann \cite{Reimann07}, and later studied in \cite{Dotsenko13};  it has recently been realised experimentally \cite{Argun17}, \alb{and extended to the underdamped regime in \cite{Mancois18}. The original Filliger and Reimann model is} based on an overdamped particle moving in a 2D anisotropic and rotated harmonic potential. 
Driven by two temperature biased heat reservoirs  the system enters a non equilibrium steady state yielding a finite torque acting on the potential. In other words, the heat transmitted to the 
system is converted into a mechanical torque giving rise to a gyrating motion, i.e., an elementary heat engine. 
This model bears a resemblance 
to a recent study in \cite{Fogedby17} where it was found that the minimal requirements for directed transport to emerge in a two-temperature autonomous 
motor are i) a non equilibrium thermal state and ii) a broken spatial symmetry.

In the present paper we construct and analyse a microscopic quantum model, fulfilling the requirements i) and ii) above,
based on an extension of the harmonic model in \cite{Reimann07}. Since a proper quantum treatment requires a Hamiltonian framework the first step is to extend
the model to the underdamped case and introduce a mass.  Moreover, the Hamiltonian description of the resulting 
open quantum system must be extended to the heat reservoirs; here modelled as collections of independent
quantum oscillators. Since the combined system is linear it is most convenient to work in the Heisenberg picture
and construct quantum Langevin equations according to the scheme first proposed by Ford, Kac, and Mazur
\cite{Ford65,Ford87,Ford88}. 

In the Ohmic approximation, corresponding to a particular spectral distribution of the oscillator modes in the heat baths, 
driving the small system in a stationary non equilibrium state, the temperature biased heat reservoirs, generate a 
finite quantum angular momentum, corresponding to a rotation. However, since the moment of inertia is fluctuating
and strongly correlated with the angular momentum, the quantum gyrator or quantum rotator does not act like a rigid
body and one  cannot meaningfully associate an angular frequency with the motion. 
At variance with the classical case, the thermal fluctuations, moreover,  give rise to an additional contribution to the average torque which is a genuinely quantum effect. Such an effect is akin to the Casimir effect for the rotatory motion, in the sense that the quantum reservoirs give rise to a quantum torque which vanishes in the classical limit.

%%%%%%%%%%%%%%%%%%%%%%%%%%%%%%%%%%%%%%%%%%%%%%%%%%%%%%%%%
%%%%%%%%%%%%%%%%%%%%%%%%%%%%%%%%%%%%%%%%%%%%%%%%%%%%%%%%%
%\section{\label{model}Quantum rotator model}
%%%%%%%%%%%%%%%%%%%%%%%%%%%%%%%%%%%%%%%%%%%%%%%%%%%%%%%%%
%%%%%%%%%%%%%%%%%%%%%%%%%%%%%%%%%%%%%%%%%%%%%%%%%%%%%%%%%
{\it Quantum rotator model}

Here we set up the quantum model for the rotator; for mathematical details we refer to appendix A. % \ref{app1}.
 We consider a particle 
of mass $m$ at position $(x_1, x_2)$ moving in a rotated 2D anisotropic harmonic potential.
The system is characterised by the quantum Hamiltonian
%\begin{eqnarray}
%&&H_0=-\frac{\hbar^2}{2m}\bigg(\frac{\partial^2}{\partial x_1^2}+\frac{\partial^2}{\partial x_1^2}\bigg)+U(x_1,x_2)
%-x_1f_1-x_2f_2,
%\label{ham1}
%\\
%&&U(x_1,x_2)=\frac{1}{2}Ax_1^2+\frac{1}{2}Bx_2^2+Cx_1x_2,
%\label{pot}
%\end{eqnarray}
\begin{eqnarray}
&&H_0=-\frac{\hbar^2}{2m}\bigg(\frac{\partial^2}{\partial x_1^2}+\frac{\partial^2}{\partial x_2^2}\bigg)+U(x_1,x_2),
\label{ham1}
\\
&&U(x_1,x_2)=\frac{1}{2}Ax_1^2+\frac{1}{2}Bx_2^2+Cx_1x_2.
\label{pot}
\end{eqnarray}
%where we have added a time dependent  external force $(f_1(t),f_2(t))$. 
The parameters $A$, $B$, and $C$ are related to the
rotated potential $u_1x_1^2+u_2x_2^2$ by
\begin{eqnarray}
&&A=u_1\cos^2\alpha+u_2\sin^2\alpha,
\label{A}
\\
&&B=u_2\cos^2\alpha+u_1\sin^2\alpha,
\label{B}
\\
&&C=\frac{1}{2}(u_1-u_2)\sin 2\alpha,
\label{C}
\end{eqnarray}
where $u_1$ and $u_2$ are the anisotropy parameters \alb{(with $u_1,\, u_2>0$ for mechanical stability)}, and  $\alpha$ the rotation angle.

The model has the structure of two linear oscillators with frequencies $\omega_1=(A/m)^{1/2}$ and
$\omega_2=(B/m)^{1/2}$ coupled linearly with strength $C$. Since $U$ originates from a rotated anisotropic
potential the parameters are constrained according to $A+B=u_1+u_2$, $A-B=(u_1-u_2)\cos 2\alpha$, and
$AB-C^2=u_1u_2$; note that the isotropic case $u_1=u_2$ implies $A=B$ and $C=0$, corresponding to
rotational invariance and two independent oscillators.

Following the prescription in \cite{Caldeira83a,Caldeira83b} the two independent heat baths including 
their coupling to the rotator coordinates are given by the quantum Hamiltonians
\begin{eqnarray}
H_n=\sum_k\bigg(\frac{P_k^{(n)2}}{2m_k^{(n)}}+
\frac{1}{2}m_k^{(n)}\omega_k^{(n)2}(X_k^{(n)}-x_n)^2\bigg),
\label{ham2}
\end{eqnarray}
with  $n=1,2$.
%\begin{eqnarray}
%H_n=\sum_k\bigg(-\frac{\hbar^2}{2m_k^{(n)}}\frac{\partial^2}{\partial X_k^{(n)2}}+
%\frac{1}{2}m_k^{(n)}\omega_k^{(n)2}(X_k^{(n)}-x_n)^2\bigg).
%\label{ham2}
%\end{eqnarray}
The heat baths are characterised by the masses $m_k^{(n)}$ and the frequencies 
$\omega_k^{(n)}$; the sum runs over the wavenumber modes $k$.
Note that the coupling to the system is absorbed in the definitions of $m_k^{(n)}$ and $\omega_k^{(n)}$
\cite{Ford88}. Introducing the density of states 
$N^{(n)}(\omega)=2\pi\sum_k m_k^{(n)}\omega_k^{(n)2}\delta(\omega-\omega_k^{(n)})$,
characterising the spectral distribution of bath nodes, the Ohmic approximation corresponds
to choosing a constant density of states, i.e.,  $N^{(n)}=2\eta$.  Non-Ohmic approximations will in general 
give rise to memory effects; they will not be considered here.

The quantum Langevin equations for a quantum rotator 
in the Ohmic approximation thus take the form
\begin{eqnarray}
m\ddot x_n=-\partial U/\partial x_n - \eta\dot x_n+\xi_n.
\label{qlan}
\end{eqnarray}
Here the quantum aspects are incorporated in the quantum operators $\xi_n$ which plays the role
of a "quantum noise". Performing a statistical quantum average we obtain the quantum relations 
$\langle\xi_n(t)\xi_m(t')\rangle=\delta_{nm} F_n(t-t')$ where
\begin{eqnarray}
 F_n(t)=
\eta\int\frac{d\omega}{2\pi}e^{-i\omega t}\hbar\omega\pq{1+\coth\Big(\frac{\hbar\omega}{2T_n}\Big)};
\label{qcor}
\end{eqnarray}
here $T_n$ are the temperatures of the respective baths and we have set $k_B=1$. For clarification we note that the quantum noise $\xi_n$ should
not be confused with a classical Gaussian coloured noise. The quantum noise originates from the equilibrium
averages over the initial bath variables. Note, however, that in the classical limit $\hbar\rightarrow 0$ the noises
commute and  we obtain $F_n(t-t')=2 \eta T_n\delta(t-t')$. As a result eq.~(\ref{qlan}) reduces to a standard 
Langevin equation driven by white Gaussian noise. The equations (\ref{qlan}) and (\ref{qcor}) form the basis
for the further analysis in the paper.

%%%%%%%%%%%%%%%%%%%%%%%%%%%%%%%%%%%%%%%%%%%%%%%%%%%%%%%%%
%%%%%%%%%%%%%%%%%%%%%%%%%%%%%%%%%%%%%%%%%%%%%%%%%%%%%%%%%
%\section{\label{angular}Quantum angular momentum and rotation}
%%%%%%%%%%%%%%%%%%%%%%%%%%%%%%%%%%%%%%%%%%%%%%%%%%%%%%%%%
%%%%%%%%%%%%%%%%%%%%%%%%%%%%%%%%%%%%%%%%%%%%%%%%%%%%%%%%%
{\it Quantum angular momentum and torque}

We here present and discuss the results for the angular momentum and torque, for mathematical details see appendix B. % \ref{app2}.
 Even in the absence of an external force, the heat reservoirs at temperatures $T_1$ and $T_2$ drive the system into
a rotating state, characterised by a non vanishing mean angular momentum.
Using the \alb{ standard definition for the angular momentum} $L=m(x_1\dot x_2-x_2\dot x_1)_W$,
\alb{where $(AB)_W=(1/2)(AB+BA)$ denotes the symmetrical Weyl ordering, see e.g. \cite{Das93}},
we find the general expression
\begin{eqnarray}
\! \! \! &&\langle L\rangle_0=-4m\eta^2C\int\frac{d\omega}{2\pi}\frac{\omega^2}{|Z(\omega)|^2}G(\omega, T_1,T_2),
\label{mom}
\\
%Z(\omega)&=&(A-m\omega^2-i\omega\eta)(B-m\omega^2-i\omega\eta)-C^2, \label{Z} \\
&&G(\omega, T_1,T_2)=\frac{\hbar\omega}{2}
\pq{\coth\Big(\frac{\hbar\omega}{2T_1}\Big)-\coth\Big(\frac{\hbar\omega}{2T_2}\Big)},
\label{G:def}
\end{eqnarray}
with $Z(\omega)=(A-m\omega^2-i\omega\eta)(B-m\omega^2-i\omega\eta)-C^2$.
We note that $\langle L\rangle_0=0$ for $C=0$, corresponding to the restoration of rotational invariance.
Moreover, $\langle L\rangle_0=0$ for $T_1=T_2$, corresponding to thermal equilibrium. In non equilibrium we have 
$\langle L\rangle_0 \propto\text{sign}(C(T_2-T_1))$.
\begin{figure}[h]
\center
%\psfrag{ }[ct][ct][1.]{ }
\includegraphics[width=8cm]{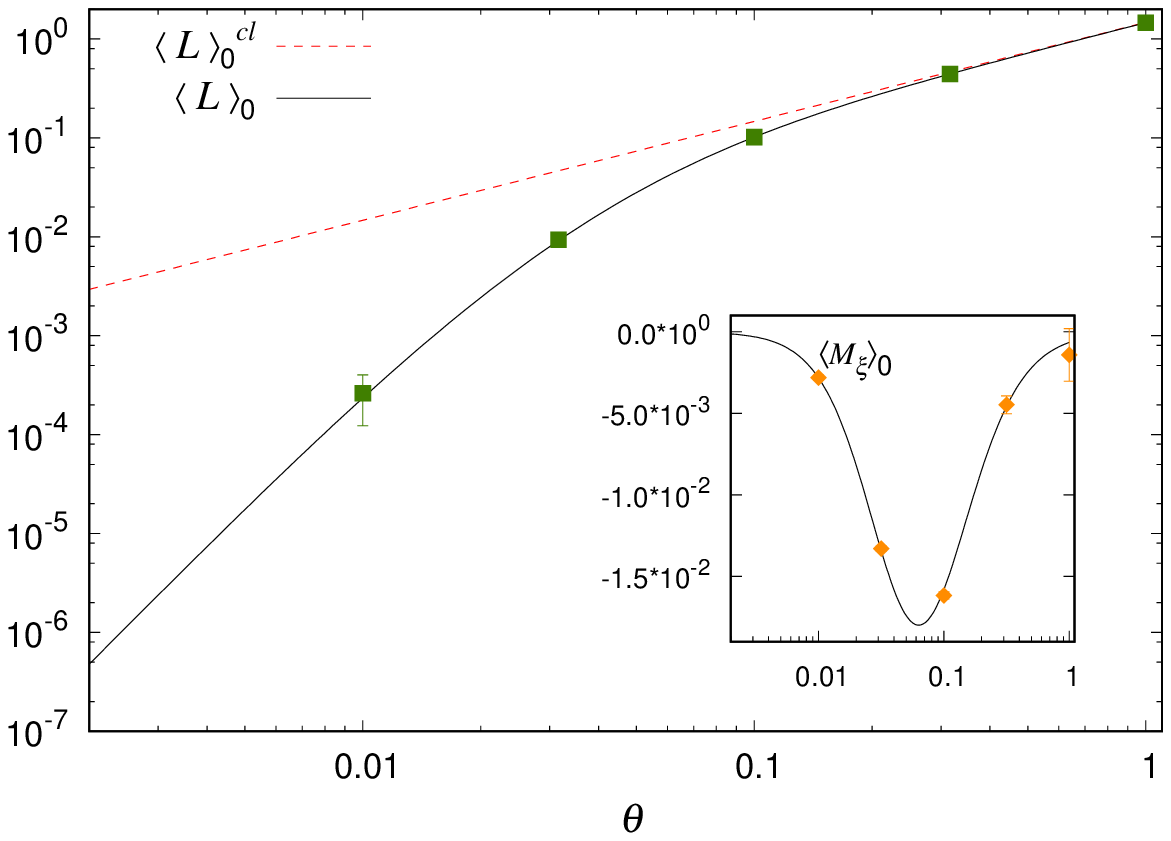}
\includegraphics[width=8cm]{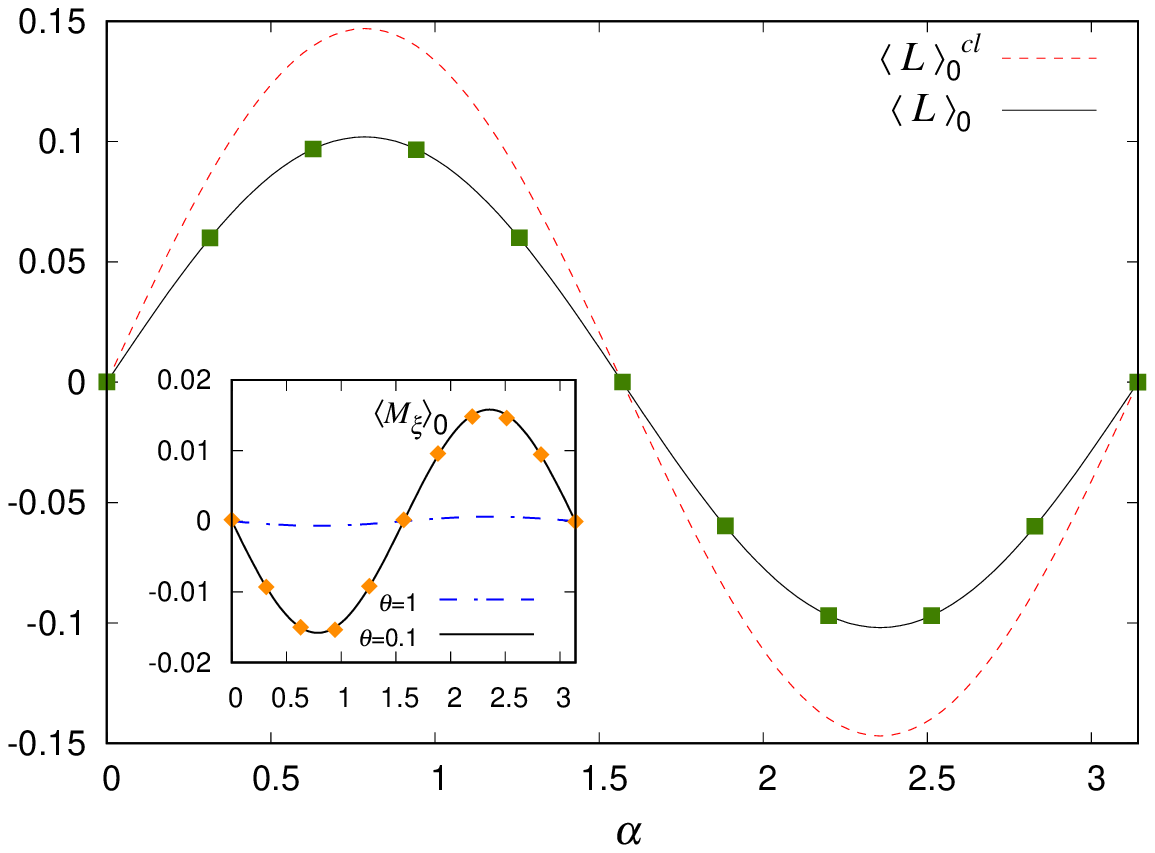}
\caption{{\it Top panel}: Quantum (eq.~(\ref{mom}), full line) and classical (eq.~(\ref{mom3}), dashed line) angular momenta as functions of the temperature scaling factor $\theta$, with  $T_1=2 \theta$, $T_2=5 \theta$, $u_1=1,\, u_2=1/4$, $\alpha=\pi/4$, $\hbar=1$, $m=\eta=1$. Inset: Quantum noise torque  $M_\xi$ as defined in eq.~(\ref{momxi}) as a function of the  temperature scaling factor 
$\theta$, the other parameters are the same as in the main figure. In both the main panel and the inset, the points are obtained from the QMD algorithm, by integrating the corresponding Langevin equation with $10^4$ independent trajectories of time duration $t_f=2^{22} \delta t$, with integration time $\delta t=10^{-3}$ in reduced units. {\it Bottom panel}: Quantum (eq.~(\ref{mom}), full line) and classical (eq.~(\ref{mom3}), dashed line) angular momenta as functions of the potential rotation angle $\alpha$, with  $T_1=0.2$, $T_2=0.5 $, $u_1=1,\, u_2=1/4$,  $m=\eta=\hbar=1$. Inset:  Quantum noise torque $M_\xi$ as a function of $\alpha$, with  $T_1=2\theta $, $T_2=5\theta $, for two different values of $\theta$,  the other parameters are the same as in the main figure.  In both the main panel and the inset, the points are obtained from the QMD algorithm. }
\label{figL}
\end{figure}

The integral in eq.~(\ref{mom}) cannot be performed analytically by contour integration in general, since the integrand has an infinite (but isolated) number of poles along the complex axis. However, in the classical 
or high temperature limit $\hbar\omega\ll T_n$,  we have $G(\omega, T_1,T_2)\rightarrow T_1-T_2$,
%$\langle L\rangle_0=-4m\eta^2C(T_1-T_2)\int(d\omega/2\pi)\omega^2/|Z(\omega)|^2$, 
and the integral in eq.~(\ref{mom}) yields
to a contour integration. In terms of the anisotropy parameters $u_1$ and $u_2$ we obtain
\begin{eqnarray}
\langle L\rangle_0^{\text{cl}}=-\frac{2m\eta(T_1-T_2)(u_1-u_2)\sin 2\alpha}{m(u_1-u_2)^2+2\eta^2(u_1+u_2)}.
\label{mom3}
\end{eqnarray}

In Fig.~\ref{figL} we have depicted $\langle L\rangle_0$ and $\langle L\rangle_0^{\text{cl}}$
as functions of the temperature scaling factor $\theta$, with $T_n=\tau_n \theta$, and $\tau_n$ constant (top panel) 
and as a function of the potential rotation angle $\alpha$  (bottom panel).

\alb{In ref.\cite{Reimann07} the classical overdamped version of the rotator was investigated and 
 the expression for the torque of the friction forces 
$\langle M_\eta\rangle_0^{\text{cl}}=\eta\langle(\dot x_1 x_2-\dot x_2 x_1)\rangle$ was obtained (here we take the two friction coefficients equal $\eta_1=\eta_2=\eta$). While the angular momentum  $L=m(x_1\dot x_2-x_2\dot x_1)$ is not defined in the overdamped limit $m\to 0$, from our result eq.~(\ref{mom3}) we can calculate the torque of the friction force as $\langle M_\eta\rangle_0^{\text{cl}}=(\eta/m)\langle L\rangle_0^{\text{cl}}$. 
By taking the limit $m\to 0$, we obtain from  eq.~(\ref{mom3})
$\langle M_\eta\rangle_0^{\text{cl}}=-(T_1-T_2)(u_1-u_2)\sin 2\alpha/(u_1+u_2)$ in accordance with the result  obtained 
in \cite{Reimann07}. 
}

Since the mean angular momentum is finite and bounded, we conclude that the total torque must vanish on average,
i.e.,  $\average{M}_0=m\average{(x_1 \ddot x_2 - x_2 \ddot x_1)_W}_0=0$. This result follows from 
$\average{M}_0=d\average{L}_0/dt$ and ergodicity and can also be checked by a direct calculation, using the solution of 
eq.~(\ref{qlan}), see appendix B. %\ref{app2}.
Consequently,  from $\average{M}_0=0$ and from eq.~(\ref{qlan}) we obtain the identity
\begin{eqnarray}
&&\frac{\eta}{m}\langle L\rangle_0=
\nonumber
\\
&&-\langle( x_1\partial_{x_2} U-x_2\partial_{x_1} U)_W\rangle_0
+\langle( x_1\xi_2-x_2\xi_1)_W\rangle_0,~~~~~
\label{identity}
\end{eqnarray}
showing that the mean angular momentum is proportional to the torque acting on the potential
plus a torque contribution from the quantum noise. 

The torque associated with the quantum noise fluctuations, 
$\average{M_\xi}_0=\langle( x_1\xi_2-x_2\xi_1)_W\rangle_0$,
is given by
\begin{eqnarray}
\average{M_\xi}_0=2C\eta\int\frac{d\omega}{2\pi}
\frac{1}{Z(\omega)}G(\omega, T_1,T_2).
\label{momxi}
\end{eqnarray}
This torque contribution vanishes when the potential is rotationally symmetric ($C=0$), at equilibrium  $T_1=T_2$,
(see the definition of $G(\omega)$ in eq.~(\ref{G:def})), and in  the classical limit $\hbar\to 0$. 
The last result can also be found  by contour integration of the integral in eq.~(\ref{momxi}), for details consult  appendix  B. % \ref{app2}.  
Since $(\eta/m)\langle L\rangle_0$ is equal to the dissipative torque $\eta \langle x_1\dot x_2-x_2\dot x_1\rangle_0$, 
the  result $\average{M_\xi}_0=0$ in the classical limit  is consistent with the analysis in the classical overdamped case in \cite{Reimann07}.
In the insets of Fig.~\ref{figL} the average quantum torque $\average{M_\xi}_0$ is plotted as a function of the temperature scaling factor $\theta$ and as a function of the angle $\alpha$.

The observation that the fluctuations in the heat baths give rise to a non vanishing torque contribution $\average{M_\xi}_0$ is a genuine quantum effect. It is akin to the Casimir effect due to quantum vacuum fluctuations, see e.g., \cite{vanEnk95,Bordag01}, 
or critical fluctuations in classical systems \cite{gambassi08}. 
In fact, the non-vanishing  stochastic force torque $\average{M_\xi}_0$ is due to the mismatch between the fluctuations in the two quantum baths, a mismatch that arises in the non-equilibrium/non-rotational invariant case.
\alb{It is worthwhile noting that the quantum noise torque  $\average{M_\xi}_0$  has always the opposite sign of the mean angular moment $\langle L\rangle_0$, for the broad range of parameter values considered in  fig.~\ref{figL}.  
Consequently, an inspection of eq.~(\ref{identity}) indicates that the non-vanishing $\average{M_\xi}_0$ decreases the value of the angular momentum thus diminishing the performance of the rotor.}

Unlike the case of a rigid body, the moment of inertia $I=m(x_1^2+x_2^2)$ for the quantum rotator is fluctuating.
This follows from the fact that the position of the particle explores the whole potential region, including the origin.
Consequently, one cannot meaningfully define an angular velocity $\Omega$ according to $L=I\Omega$.
In other words, there are correlations between $I$ and $\Omega$ and the approximation $\langle L\rangle\sim\langle I\rangle\langle\Omega\rangle$ is not a priori justified. Based on a numerical integration of the Langevin equations
in the classical limit reported in appendix B.1 %\ref{app2}.1, % \ref{app2:angvel}, 
we have demonstrated that the moment of inertia and the angular momentum are in fact strongly correlated.
%, we have in Fig. 2 depicted $\langle L/m(x^2_1+x_2^2)\rangle$ and 
%$\langle L\rangle/m\langle(x^2_1+x_2^2)\rangle$ showing that the moment of inertia and the angular momentum are
%strongly correlated.

%%%%%%%%%%%%%%%%%%%%%%%%%%%%%%%%%%%%%%%%%%%%%%%%%%%%%%%%%
%%%%%%%%%%%%%%%%%%%%%%%%%%%%%%%%%%%%%%%%%%%%%%%%%%%%%%%%%
%\section{\label{driven}Driven quantum rotator}
%%%%%%%%%%%%%%%%%%%%%%%%%%%%%%%%%%%%%%%%%%%%%%%%%%%%%%%%%
%%%%%%%%%%%%%%%%%%%%%%%%%%%%%%%%%%%%%%%%%%%%%%%%%%%%%%%%%
{\it Driven quantum rotator}

We next consider the case where an external time dependent force, ${\boldsymbol f}(t)=(f_1(t),f_2(t))$, is applied to the system. 
In this case the Hamiltonian in eq.~(\ref{ham1}) takes the form
$H_0-\boldsymbol f\cdot \boldsymbol x$; details of this section are  given in appendix  B.2 and C. %\ref{app2}.2. and \ref{app3}. 

Applying an external periodic force the quantum oscillator is driven into a periodic state, i.e., a limit cycle.
Here we drive the quantum rotator with a periodic drive with amplitude $D$ and frequency $\omega_0$, setting
${\boldsymbol f}(t)=D(\cos\omega_0t,\sin\omega_0t)$. A time-periodic protocol is a standard choice in the implementation of microscopic cyclic devices, both in the classical and in the quantum regime, see, e.g., \cite{Imparato08,Koski2013}.
%
%\begin{eqnarray}
%&&f_1(t)=D\cos\omega_0t,
%\label{f1}
%\\
%&&f_2(t)=D\sin\omega_0t.
%\label{f2}
%\end{eqnarray}
%
For the angular momentum we obtain in the steady state the expression
\begin{eqnarray}
%\langle L\rangle&=&\langle L\rangle_0+\langle L(\omega_0)\rangle_D,\nonumber\\
%\langle L(\omega_0)\rangle_D&=&
%mD^2\omega_0\frac{(A-m\omega_0^2)(B-m\omega_0^2)+(\omega_0\eta)^2-C^2}{|Z(\omega_0)|^2}.\nonumber
\langle L\rangle&=&\langle L\rangle_0+D^2\omega_0 K(\omega_0),
\\
K(\omega_0)&=&m\frac{(A-m\omega_0^2)(B-m\omega_0^2)+\omega_0^2\eta^2-C^2}
{|Z(\omega_0)|^2}.~~~~.
\label{drmom}
\end{eqnarray}
The angular momentum is composed of two parts. The unperturbed angular momentum $\langle L\rangle_0$
generated by the broken symmetry and the temperature bias plus a contribution $D^2\omega_0 K(\omega_0)$ due to the drive. 
We note that the contribution from the drive is odd in the applied frequency $\omega_0$.
Consequently, choosing a frequency $\omega_0$ so that $\omega_0=-\langle L\rangle_0/D^2K(\omega_0)$
we can on the average arrest the rotary motion.

%%%%%%%%%%%%%%%%%%%%%%%%%%%%%%%%%%%%%%%%%%%%%%%%%%%%%%%%%
%%%%%%%%%%%%%%%%%%%%%%%%%%%%%%%%%%%%%%%%%%%%%%%%%%%%%%%%%
%\section{\label{heat}Heat and work}
%%%%%%%%%%%%%%%%%%%%%%%%%%%%%%%%%%%%%%%%%%%%%%%%%%%%%%%%%
%%%%%%%%%%%%%%%%%%%%%%%%%%%%%%%%%%%%%%%%%%%%%%%%%%%%%%%%%
{\it Work and heat }

Finally, we consider the thermodynamic properties of the driven system.
The fluctuating rate of work, $r_w$, performed on the system by the external drive is
\begin{eqnarray}
r_w=-x_1\dot f_1-x_2\dot f_2.
\label{work}
\end{eqnarray}
By insertion of the solutions of eq.~(\ref{qlan}) and averaging we obtain for the periodic drive the mean rate of work
\begin{eqnarray}
\langle r_w\rangle=&&\frac{\eta(D\omega_0)^2}{2|Z(\omega_0)|^2}\times
\nonumber
\\
&&\big [(A-m\omega_0^2)^2
+(B-m\omega_0^2)^2+2(\omega_0\eta)^2+ 2C^2\big].~~~~
\label{mwork}
\end{eqnarray}
We note that the work performed on the system is always positive, indicating that the system cannot perform work
on the environment and thus cannot perform as an engine or a motor \alb{for this specific manipulation protocol}.

It follows from eq.~(\ref{qlan}) that the fluctuating forces associated with the heat reservoirs are given
by $(-\eta\dot x_n(t)+\xi_n(t))$. For the rates of heat acting on the rotator we then have
$r_{q_n}=(\dot x_n(-\eta\dot x_n(t)+\xi_n(t)))_W$. Inserting the equations of motion we obtain for the fluctuating rates
\begin{eqnarray}
&&r_{q_1}=(\dot x_1(m\ddot x_1 +A x_1+C x_2-f_1))_W,
\label{heat1}
\\
&&r_{q_2}=(\dot x_2(m\ddot x_2 +B x_2+C x_1-f_2))_W.
\label{heat2}
\end{eqnarray}

The non equilibrium heat transfer rate is given by $\langle\Delta r_q\rangle=\langle r_{q_1}\rangle-\langle r_{q_2}\rangle$.

From the definition $\langle L\rangle=m\langle(x_1\dot x_2-x_2\dot x_1)_W\rangle$ and the definitions in 
eqs.~(\ref{heat1}) and (\ref{heat2}) we obtain the identity
\begin{eqnarray}
\langle\Delta r_q\rangle=-\frac{C}{m}\langle L\rangle-\langle\dot x_1\rangle f_1+\langle\dot x_2\rangle f_2.
\label{identity2}
\end{eqnarray}
In the absence of a drive we obtain in particular
%$\langle\Delta r_q\rangle_0=-C/m\,  \langle L\rangle_0$.
%
\begin{eqnarray}
\langle\Delta r_q\rangle_0=-\frac{C}{m}\langle L\rangle_0.
\label{identity3}
\end{eqnarray}
showing that the absorption of heat is completely converted into rotation.

\begin{figure}[h]
\center
\psfrag{ }[ct][ct][1.]{ }
\includegraphics[width=8cm]{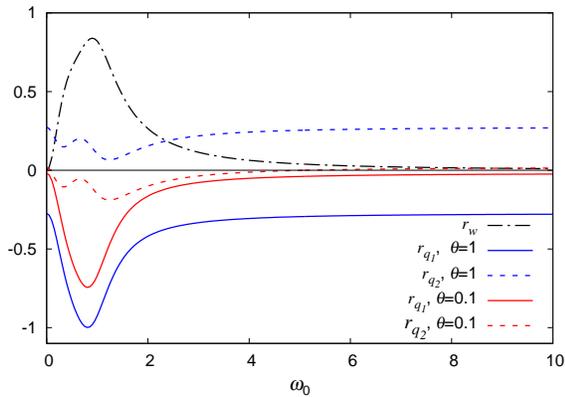}
\caption{Work rate $r_w$, and heat rates $r_{q_1}$, and $r_{q_2}$ as functions of the driving frequency $\omega_0$ as given by eqs.~(\ref{mwork}), ~(\ref{mq1}), (\ref{mq2}), respectively. The curves are obtained with the following set of parameters:    $T_1=2 \theta$, $T_2=5 \theta$, $u_1=1,\, u_2=1/4$, $\alpha=\pi/4$, $\hbar=1$, $m=\eta=D=1$}
\label{therm:fig}
\end{figure}

%In the case of the periodic drive considered above we find
%
%\begin{eqnarray}
%\langle\Delta  q\rangle&=&-\frac{C}{m}\langle L\rangle_0\nonumber\\
%&&+\frac{\eta}{2}(D\omega_0)^2\frac{(A-m\omega_0^2)^2-(B-m\omega_0^2)^2-4C(\eta\omega_0)}
%{|Z(\omega_0)|^2},\nonumber
%\label{mdelq}
%\end{eqnarray}
%
%composed of a part proportional to the intrinsic angular momentum and a part due to the drive.
For the individual heat transfers $r_{q_1}$ and $r_{q_2}$ we find 
\begin{eqnarray}
\langle r_{q_1}\rangle=&&-\frac{C}{2m}\langle L\rangle_0
\nonumber
\\
&&-\frac{\eta}{2}(D\omega_0)^2
\frac{(B-m\omega_0^2)^2+(C+\omega_0\eta)^2}{|Z(\omega_0)|^2},
\label{mq1}
\\
\langle r_{q_2}\rangle=&&+\frac{C}{2m}\langle L\rangle_0
\nonumber
\\
&&-\frac{\eta}{2}(D\omega_0)^2
\frac{(A-m\omega_0^2)^2+(C-\omega_0\eta)^2}{|Z(\omega_0)|^2},
\label{mq2}
\end{eqnarray}
each composed of a part proportional to the intrinsic angular momentum and a part due to the drive.
Assuming $T_2>T_1$ and noting that $C\langle L\rangle_0\propto\text{sign}(T_2-T_1)$, see eq.~(\ref{mom}), we infer that the average heat
transfer $\average{r_{q_1}}$ from the cold reservoir at temperature $T_1$ is always negative.  \alb{The power injected by the external driving is thus dissipated in the heath baths with rates which are given by the additional terms in eqs.~(\ref{mq1})--(\ref{mq2}) and which depend on the drive.}
 Consequently, the system
cannot work as a heat pump transferring heat from the cold to the hot reservoir.
The absence of both motor and pump performance for this manipulation protocol is in agreement with the 
analysis in \cite{Marathe07}, where it was demonstrated that a driven system composed of  
two classical linear or non-linear oscillators fails to perform either as a pump or as an engine. 
%The impossibility of building a refrigerator out of a network of quantum linear oscillator was also established in \cite{Martinez2013}.
Finally, we notice that our results for $\langle r_{q_1}\rangle, \, \langle r_{q_2}\rangle$ and $\langle r_w\rangle$  are compatible with energy conservation, i.e., $\langle r_{q_1}\rangle+\langle r_{q_2}\rangle=-\langle r_w\rangle$.  
An example of the work and heat rates, for a given choice of the system parameters, is shown in fig.~\ref{therm:fig}.

{\it Numerical algorithm}

Our expressions for the average angular momentum in eq.~(\ref{mom}) and the quantum torque in eq.~(\ref{momxi}) are exact results for two non-linear quantities in an out-of-equilibrium system. They are thus excellent test beds for checking  the validity of  the 
Quantum Molecular Dynamics (QMD) numerical  algorithm \cite{Dammak09} in the non-equilibrium regime. This algorithm accounts for quantum statistics while using standard molecular dynamics. It has been tested  and  has provided accurate results, for example, for
the average energy and the specific heat in different types of systems at equilibrium with  various degrees of anharmonicity \cite{Dammak09,Dammak12,Calvo12,Calvo12a,Qi12,Bronstein14}.
In the QMD algorithm the Heisenberg equations of motion for the quantum coordinate operator $x$, as given by eq.~(\ref{qlan}), 
is replaced by a classical Langevin equation with Gaussian coloured noise $\xi(t)$, 
whose  power spectral density  is dictated by the quantum mechanical fluctuation-dissipation relation, as expressed 
by eq.~(\ref{qcor}).
\alb{In the standard Langevin equation for classical systems the total time $t$ of a single trajectory is divided into $N_t=t/\delta t$ time steps of length $\delta t$ and at each time step the Gaussian white noise is generated through a random generator. The QMD exhibits an additional complication with respect to this scheme and is thus  more computationally demanding, i.e.,  at the beginning of each trajectory one has to generate and store an $N_t$-long vector of correlated random forces with correlation function  given by eq.~(\ref{qcor}); this is the scheme discussed in \cite{Dammak09}.}

While the QMD algorithm has previously been used in order to study the interaction of a quantum system with a single bath, we here implement it for the two-bath rotator. The results for the average 
angular momentum $\average{L}_0$ and the quantum noise torque $\average{M_\xi}$ are reported in Fig.~\ref{figL}. The agreement with our exact results is excellent.

{\it Conclusions}

In conclusion, we have studied a minimal model of a quantum thermal motor exhibiting a non-vanishing angular momentum when moving in an asymmetric and rotated potential, while interacting with two quantum heat baths at different temperatures. \alb{The rotational motion  is sustained by the heat current flowing through the system}.
There is a striking difference between the classical and the quantum rotator, in the latter case the bath forces give rise to a systematic torque contribution which is absent in the classical case. This effect is analogous to the Casimir effect.
%Indeed, the the Casimir force between two solid  plates was calculated by Lifshitz \cite{Lifshitz56} by adding a "random" field to the Maxwell equations, with a correlation function analogous to our eq.~(\ref{qcor}).

\alb{The rotor cannot perform useful work when manipulated with an external periodic load. This conclusion agrees with previous studies of classical motors where the working fluids are coupled oscillators at different temperatures. However, different designs where the trapping potential is non-linear or where the rotor is connected to a work repository, as in \cite{Levy16}, could change this conclusion and yield a device from which useful work can be extracted.} 

Since we have exact results for the average values of the systems dynamic quantities, we also checked whether an ab-initio numerical algorithm for equilibrium quantum simulations could be used in order to investigate the dynamics in the case of multiple baths. We find an excellent agreement with our results.  This result thus paves  the way to the algorithm application to the study of the dynamics in the out-of-equilibrium regime. 
\alb{In particular, since the equilibrium QMD algorithm has been successfully used on non-linear systems, it is of interest to use the non-equilibrium QMD algorithm in order to investigate non-linear out-of-equilibrium quantum systems  
for which there is no analytic expression for the dynamic and thermodynamic quantities of interest.}

%\bibliography{/Users/hansfogedby/Library/texmf/bibtex/bib/articles,books}

\begin{acknowledgments}
A.I. was supported by the Danish Council for Independent Research and the Villum Foundation.
The numerical  results presented in this work were obtained at the Centre for Scientific Computing, Aarhus http://phys.au.dk/forskning/cscaa.
\end{acknowledgments}
\def\url#1{}
\bibliography{bibliography}
\newpage

\begin{widetext}
\appendix

\section{\label{app1}Quantisation and solution}
%%%%%%%%%%%%%%%%%%%%%%%%%
The quantum rotator constitutes an open quantum system \cite{Breuer02} where a small quantum system interacts
with its surroundings; here two heat baths maintained at different temperatures. Unlike the classical case
where a separation of time scales leads to the standard description of a small system in terms of Langevin equations,
alternatively, Fokker-Planck or Master equations, the quantum case requires a complete quantum treatment
in terms of a Hamiltonian for the whole system. Following \cite{Caldeira83a} we describe the heat reservoirs as
a collection of independent quantum oscillators. Introducing the momentum variables
\begin{eqnarray}
&&p_n=-i\hbar\frac{\partial}{\partial x_n},
\label{apn}
\\
&&P_k^{(n)}=-i\hbar\frac{\partial}{\partial X_k^{(n)}},
\label{aPkn}
\end{eqnarray}
the Hamiltonians given in eqs.~(\ref{ham1}) and (\ref{ham2}) then take the form
\begin{eqnarray}
&&H=\frac{1}{2m}(p_1^2+p_2^2)+U(x_1,x_2)
-x_1f_1-x_2f_2,
\label{aham1}
\\
&&U(x_1,x_2)=\frac{1}{2}Ax_1^2+\frac{1}{2}Bx_2^2+Cx_1x_2,
\label{apot}
\\
&&H_n=\sum_k\bigg(\frac{P_k^{(n)2}}{2m_k^{(n)} }+
\frac{1}{2}m_k^{(n)}\omega_k^{(n)2}(X_k^{(n)}-x_n)^2\bigg),~~~n=1,2.
\label{aham2}
\end{eqnarray}
where we have already included the external force ${\boldsymbol f}$.
Since the total system is linear a complete quantisation is most conveniently carried out in the Heisenberg picture 
following the scheme proposed by Ford, Kac, and Mazur \cite{Ford65}. 

Applying the Heisenberg  equation of motion, $i\hbar dA/dt=[A,H]$, to the Hamiltonians $H$ and $H_n$,
$n=1,2$ in eqs.~(\ref{aham1}) and (\ref{aham2}) we obtain
\begin{eqnarray}
&&m\ddot x_n+\frac{\partial U}{\partial x_n}=\sum_k m_k^{(n)}\omega_k^{(n)2}(X_k^{(n)}-x_n)+f_n,
\label{aeq1}
\\
&&\ddot X_k^{(n)}+\omega_k^{(n)2}X_k^{(n)}=\omega_k^{(n)2}x_n.
\label{aeq2}
\end{eqnarray}
Solving first the equations for the bath variables $X_k^{(n)}$ we have
\begin{eqnarray}
X_k^{(n)}(t)&=&X_k^{(n)}\cos\omega_k^{(n)}(t-t_0)
+\dot X_k^{(n)}\frac{\sin\omega_k^{(n)}(t-t_0)}{\omega_k^{(n)}}
\nonumber
\\
&+&x_n(t)-x_n\cos\omega_k^{(n)}(t-t_0)
-\int_{t_0}^t dt'\dot x_n(t')\cos\omega_k^{(n)}(t-t'),
\label{asolX}
\end{eqnarray}
where $X_k^{(n)}=X_k^{(n)}(t_0)$, $\dot X_k^{(n)}=\dot X_k^{(n)}(t_0)$, and $x_n=x_n(t_0)$ are the initial values at time $t=t_0$.  
Next, inserting the solution for the bath variables in the equations for $x_n$ we obtain
\begin{eqnarray}
m\ddot x_n(t)+\int_{t_0}dt'\eta_n(t-t')\dot x_n(t')+\frac{\partial U}{\partial x_n(t)}+x_n\eta^{(n)}(t)=\xi_n(t)+f_n,
\label{aeq11}
\end{eqnarray}
where, introducing the step function  $\theta(t)$, we have
\begin{eqnarray}
&&\eta_n(t)=\theta(t)\sum_k m_k^{(n)}\omega_k^{(n)2}\cos\omega_k^{(n)} t,
\label{aeta}
\\
&&\xi_n(t)=\sum_k m_k^{(n)}\omega_k^{(n)2}\bigg[X_k^{(n)}\cos\omega_k^{(n)}(t-t_0)+\dot X_k^{(n)}
\frac{\sin\omega_k^{(n)}(t-t_0)}{\omega_k^{(n)}}\bigg].
\label{axi1}
\end{eqnarray}
Here eq.~(\ref{aeq11}) already has the structure of a general Langevin equation. The time dependent damping
is given by $\eta_n(t)$, whereas all reference to the reservoirs is contained in the quantum noise $\xi_n(t)$. 
Assuming that the system is uncoupled prior to $t=t_0$ and that the bath variables $X_k^{(n)}$
are in equilibrium at temperatures $T_n$ we average over $X_k^{(n)}$ and $\dot X_k^{(n)}$
according to
\begin{eqnarray}
&&\langle X_k^{(n)}X_p^{(m)}\rangle=\delta_{kp}\delta_{nm}\bigg(\frac{\hbar}{2m_k^{(n)}\omega_k^{(n)}}\bigg)
\coth\frac{\hbar\omega_k^{(n)}}{2T_n},
\label{aprod1}
\\
&&\langle\dot X_k^{(n)}\dot X_p^{(m)}\rangle=\delta_{kp}\delta_{nm}\bigg(\frac{\hbar\omega_k^{(n)}}{2m_k^{(n)}}\bigg)
\coth\frac{\hbar\omega_k^{(n)}}{2T_n},
\label{aprod2}
\\
&&\langle X_k^{(n)}\dot X_p^{(m)}\rangle=\delta_{kp}\delta_{nm}\bigg(\frac{i\hbar}{2m_k^{(n)}}\bigg).
\label{aprod3}
\end{eqnarray}
Introducing the density of states
\begin{eqnarray}
N^{(n)}(\omega)=2\pi\sum_k m_k^{(n)}\omega_k^{(n)2}\delta(\omega-\omega_k^{(n)}),
\label{aDOS}
\end{eqnarray}
we have
\begin{eqnarray}
&&\eta_n(t)=\theta(t)\int\frac{d\omega}{2\pi} N^{(n)}(\omega)\cos\omega t,
\label{aeta1}
\end{eqnarray}
and for the correlations of the quantum noise
\begin{eqnarray}
\langle\xi_n(t)\xi_m(t')\rangle=\frac{1}{2}
\delta_{nm}\int\frac{d\omega}{2\pi}N^{(n)}(\omega)\hbar\omega
\bigg[\cos\omega(t-t')\coth\frac{\hbar\omega}{2T_n}
-i\sin\omega(t-t')\bigg].
\label{qcorr1}
\end{eqnarray}
The commutator and anti commutators are thus given by
\begin{eqnarray}
&&[\xi_n(t),\xi_m(t')]=-i\delta_{nm}\int\frac{d\omega}{2\pi} N^{(n)}(\omega)\hbar\omega\sin\omega(t-t'),
\label{acom}
\\
&&\{ \xi_n(t),\xi_m(t')   \}=
\delta_{nm}\int\frac{d\omega}{2\pi}N^{(n)}(\omega)\hbar\omega\cos\omega(t-t')\coth\frac{\hbar\omega}{2T_n}.
\label{aanticom}
\end{eqnarray}
In the Ohmic approximation the density of states is assumed to be constant. Setting $N^{(n)}=2\eta$
we obtain the quantum Langevin equations
\begin{eqnarray}
m\ddot x_n=-\frac{dU}{dx_n} - \eta\dot x_n+f_n+\xi_n.
\label{aqlan}
\end{eqnarray}
with noise correlations
\begin{eqnarray}
\langle\xi_n(t)\xi_m(t')\rangle=\delta_{nm}
\eta\int\frac{d\omega}{2\pi}e^{-i\omega(t-t')}\hbar\omega(1+\coth(\hbar\omega/2T_n)).
\label{aqcor1}
\end{eqnarray}

For the solution of eq.~(\ref{aqlan}) we have, introducing  $x_n(t)=\int (d\omega/2\pi)\exp(-i\omega t)\tilde x_n(\omega)$, etc.,
and expanding the
potential in eq.~(\ref{apot}), the coupled Langevin equations 
\begin{eqnarray}
&&(A-m\omega^2-i\omega\eta)\tilde x_1+C\tilde x_2=\tilde\xi_1+\tilde f_1,
\label{alan11}
\\
&&(B-m\omega^2-i\omega\eta)\tilde x_2+C\tilde x_1=\tilde\xi_2+\tilde f_2,
\label{alan22}
\end{eqnarray}
with Green's function solutions
\begin{eqnarray}
&&x_1(t)=\int K_1(t-t')(\xi_1(t')+f_1(t'))dt'+\int K_2(t-t')(\xi_2(t')+f_2(t'))dt',
\label{asol1}
\\
&&x_2(t)=\int L_1(t-t')(\xi_1(t')+f_1(t'))dt'+\int L_2(t-t')(\xi_2(t')+f_2(t'))dt'.
\label{asol2}
\end{eqnarray}
Here the Fourier transforms of $K_n$ and $L_n$ are given by
\begin{eqnarray}
&&\tilde K_1(\omega)=\frac{B-m\omega^2-i\omega\eta}{Z(\omega)}\label{K1:def},
\
\\
&&\tilde K_2(\omega)=\frac{-C}{Z(\omega)}\label{K2:def},
\\
&&\tilde L_1(\omega)=\frac{-C}{Z(\omega)}\label{L1:def},
\\
&&\tilde L_2(\omega)=\frac{A-m\omega^2-i\omega\eta}{Z(\omega)}\label{L2:def},
\\
&&Z(\omega)=(A-m\omega^2-i\omega\eta)(B-m\omega^2-i\omega\eta)-C^2.
\end{eqnarray}
The above solutions form the basis for the remaining analysis.
%%%%%%%%%%%%%%%%%%%%%%%%%
\section{\label{app2} Angular momentum and torque}
%%%%%%%%%%%%%%%%%%%%%%%%%
We first consider the angular momentum in the absence of an external force (${\boldsymbol f}=0$), i.e., the intrinsic
angular momentum. In 2D the fluctuating angular momentum is defined as
\begin{eqnarray}
L(t)=m(x_1(t)\dot x_2(t)- x_2(t)\dot x_1(t))_W;
\label{defmom}
\end{eqnarray}
note that $(AB)_W=(1/2)(AB+BA)$ indicates the symmetric Weyl ordering,
see e.g., \cite{Das93}.
For the average intrinsic angular momentum $\langle L\rangle_0=
m\langle(x_1\dot x_2-x_2\dot x_1)_W\rangle$
we obtain by insertion of eqs.~(\ref{asol1}) and (\ref{asol2})
and averaging over the noise  according to eq.~(\ref{aqcor1})
\begin{eqnarray}
&&\langle L\rangle_0=-4m\eta^2C\int\frac{d\omega}{2\pi}\frac{\omega^2}{|Z(\omega)|^2}G(\omega, T_1,T_2)
\label{aL0},
\label{mom11}
\\
&&G(\omega, T_1,T_2)=\frac{\hbar\omega}{2}(\coth(\hbar\omega/2T_1)-\coth(\hbar\omega/2T_2)).
\label{aG}
\end{eqnarray}
In the classical limit for $\hbar\rightarrow 0$,
expanding $G$, we obtain
\begin{eqnarray}
\langle L\rangle_0^{\text{cl}}=-4m\eta^2C(T_1-T_2)\int\frac{d\omega}{2\pi}\frac{\omega^2}{|Z(\omega)|^2}.
\label{aLcl}
\end{eqnarray}
In terms of $u_1$ and $u_2$, using eqs.~(\ref{A}-\ref{C}) in the main text,  one obtains
$Z(\omega)=(u_1^2-\omega^2(i \eta -m \omega)^2)(u_2^2-\omega^2(i \eta -m \omega)^2)$
 with zeros
$\omega_{1,\pm}=(i\eta\pm\sqrt{4mu_1-\eta^2})/2m$ and $\omega_{2,\pm}=(i\eta\pm\sqrt{4mu_2-\eta^2})/2m$, and the corresponding conjugate points.
Consequently, the integrand in eq.~(\ref{aLcl})  have four poles in the upper half plane and four 
conjugate poles in the lower half plane. By contour integration we obtain
\begin{eqnarray}
\langle L\rangle_0^{\text{cl}}=-\frac{2m\eta(T_1-T_2)(u_1-u_2)\sin 2\alpha}{m(u_1-u_2)^2+2\eta^2(u_1+u_2)}.
\end{eqnarray}
In addition to $\omega_{1,\pm}$ and $\omega_{2,\pm}$, the integrand of eq.~(\ref{aL0}) has an infinite (but isolated) number of poles along the positive complex axis $\omega_{n,p}=2\pi i p T_n /\hbar$,  $n=1,2$ and $p=1,2,3,\dots$. As such it can be calculated by, e.g., Mathematica, as the summation of an infinite series of residues. 

Since the mean angular momentum  $\langle L(t)\rangle_0$ is finite and bounded 
it follows from $M=dL/dt$ \cite{Landau59b} and ergodicity that $\langle M\rangle_0=0$; this is also
the result of a direct calculation using the definition $M=m(x_1\ddot x_2-x_2\ddot x_1)_W$ and the solutions
in appendix \ref{app1}. From the equations of motion (\ref{aqlan}) we then obtain the identity
\begin{eqnarray}
-\langle(x_1\partial U/\partial x_2-x_2\partial U/\partial x_1)_W\rangle_0-\frac{ \eta}{m}\langle L\rangle_0+\langle(x\xi_2-y\xi_1)_W\rangle_0=0.
\end{eqnarray}

The noise torque is given by
\begin{eqnarray}
\langle M_\xi\rangle_0=\langle(x_1\xi_2-x_2\xi_1)_W\rangle_0.
\label{noise_t:def}
\end{eqnarray}
In the classical limit we find by insertion of the solutions in eqs.~(\ref{asol1}) and (\ref{asol2})
and averaging over the noise  according to eq.~(\ref{aqcor1})
\begin{eqnarray}
\langle M_\xi\rangle^{cl}_0 = 2 \eta(T_1-T_2)C\int\frac{d\omega}{2\pi}\frac{1}
{Z(\omega)}.
\end{eqnarray}
By inspection the integrand only have poles in the lower half plane and closing the contour in
the upper half plane we obtain $\langle M_\xi\rangle=0$. 
However, surprisingly, this result does not hold in the quantum regime where we obtain 
$\langle M_\xi\rangle_0\neq 0$. A simple calculation in fact yields
\begin{eqnarray}
\langle M_\xi\rangle_0 = 
 C \eta\int\frac{d\omega}{2\pi}\frac{\hbar\omega}
{Z(\omega)}\pq{\coth(\hbar\omega/2T_1)-\coth(\hbar\omega/2T_2)},
\label{M0}
\end{eqnarray}
which has poles $\omega_{n,p}$ in the upper half plane.
Thus we conclude that the quantum noise torque is non vanishing in the non-equilibrium/non-rotational invariant case ($T_1\neq T_2$ and $C\neq 0$).
Similarly to $\average{L}_0$, the  noise torque  $\langle M_\xi\rangle_0$ can be calculated as the summation of an infinite series of residues, by using, e.g., Mathematica. Differently from $\average{L}_0$ its dependence on the sign of $(T_1-T_2)$ cannot be determined analytically. Nevertheless, for the broad range of parameters considered in this manuscript, we find that  $\average{L}_0$ and  $\langle M_\xi\rangle_0$ have opposite sign, see fig.~\ref{figL} in the main text.

%\alb{In order to evaluate the integral in eq.~(\ref{M0}), one has to evaluate the residues of the function $g(z,T_n)=C\eta z \coth(\hbar z/2 T_n)Z^{-1}(z)/2 \pi$ at the poles $z_p=2 p i \pi T_n/\hbar$, $p=1,2,\dots$ .
%One finds
%\begin{equation}
%2 \pi i \mathrm{Res}[g(z_p,T_n)]=\frac{-4  C\eta  \pi  \hbar^3 p T_n^2}{\left(\hbar^2 u_1+2 \pi  p T_n (\eta  \hbar+2 \pi  p m T_n)\right) \left(\hbar^2 u_2+2 \pi  p T_n (\eta  \hbar+2 \pi  p m T_n)\right)}.
%\label{Res}
%\end{equation} 
%The integral in  eq.~(\ref{M0}) is thus equal to the sum of the series of coefficients
%\begin{equation}
%c_p=2 \pi i (\mathrm{Res}[g(z_p,T_1)]-\mathrm{Res}[g(z_p,T_2)])
%\end{equation} 
%and by noticing that all the parameters appearing in eq.~(\ref{Res}) are positive, one easily concludes that $c_k\propto \text{sign}(C(T_1-T_2))$ and thus $\langle M_\xi\rangle_0\propto\text{sign}(C(T_1-T_2))$.
%}
%\hf{Problem!!!! for $T_1>T_2$ the quantity $c_p$ is definitely positive for large enough $p$, for small $p$ it can be negative}
%%%%%%%%%%%%%%
\subsection{Angular velocity}
\label{app2:angvel}
%%%%%%%%%%%%%%
Both the classical and the quantum rotator exhibit a finite angular momentum corresponding to rotation. 
However, it does not seem possible to identify a proper angular frequency $\Omega$ since the moment of inertia
$I$, unlike the case of a rigid body \cite{Landau59b}, is fluctuating and correlated with the angular momentum.
In more detail, noting that the fluctuating angular momentum is given by eq.~(\ref{defmom})
and introducing polar coordinates, $x_1(t)=r(t)\cos\theta(t),~x_2(t)=r(t)\sin\theta(t)$, we have
\begin{eqnarray}
&&L(t)=I(t)\Omega(t),
\\
&&\Omega(t)=\dot \theta(t),
\\
&&I(t)=m r(t)^2,
\\
&&r(t)^2=x_1^2(t)+x_2^2(t).
\end{eqnarray}
Here $\Omega (t)$ is the angular frequency and $r(t)$ the radius vector.
In order to check the degree of "rigid body" behaviour which would have allowed a definition
of an average angular velocity according to the relationship
$\average{L}_0\sim\average{I}_0\average{\Omega}_0$, we have considered the classical underdamped rotator 
and integrated the Langevin equations in order to evaluate $\average {L/r^2}_0$, and individually $\average {L}_0$ and $\average{r^2}_0$ for a set of model parameters. We find that $\average {L/r^2}_0\neq \average {L}_0/\average{r^2}_0$, see Fig.~\ref{fig:check}, corroborating the absence of a proper mean angular frequency
\begin{figure}[h]
\center
\psfrag{Lav}[rc][rc][.8]{$\average{L}$}
\psfrag{\alpha}[ct][ct][1.]{$\alpha$}
\psfrag{T2}[ct][ct][1.]{$T_2$}
\includegraphics[width=8cm]{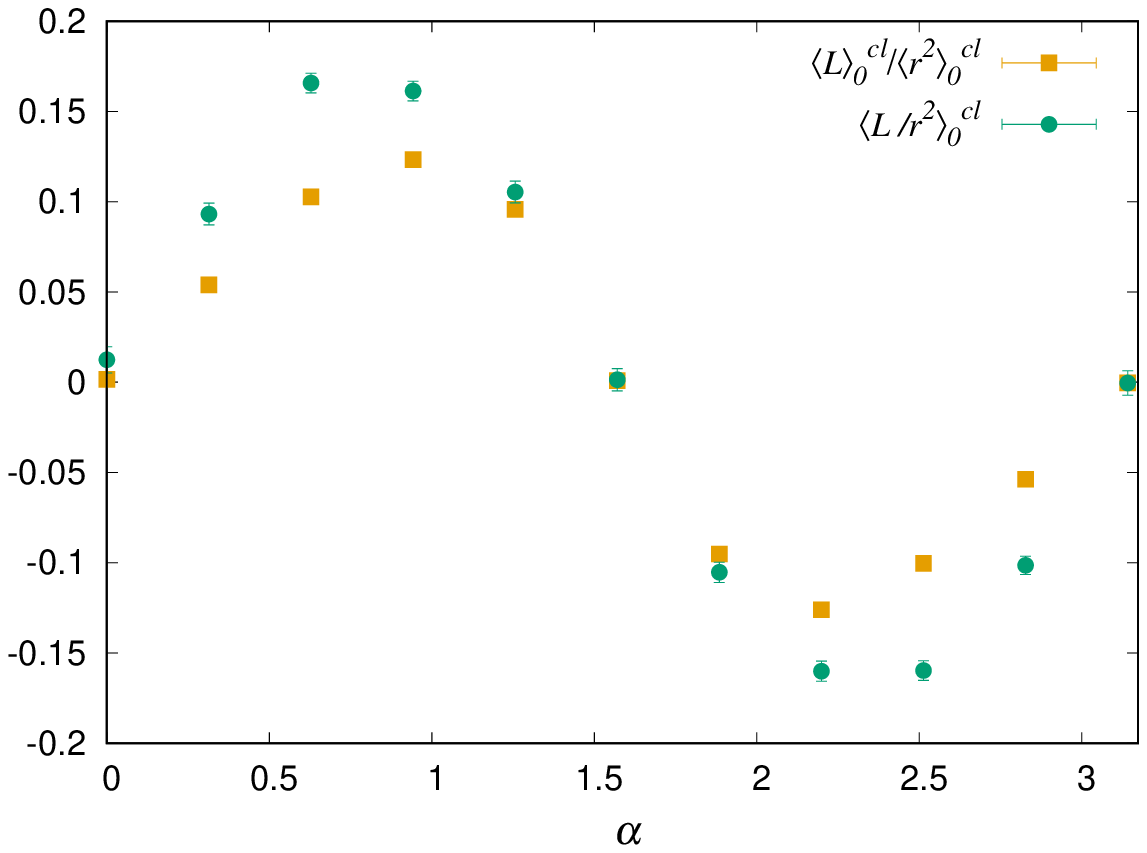}
\includegraphics[width=8cm]{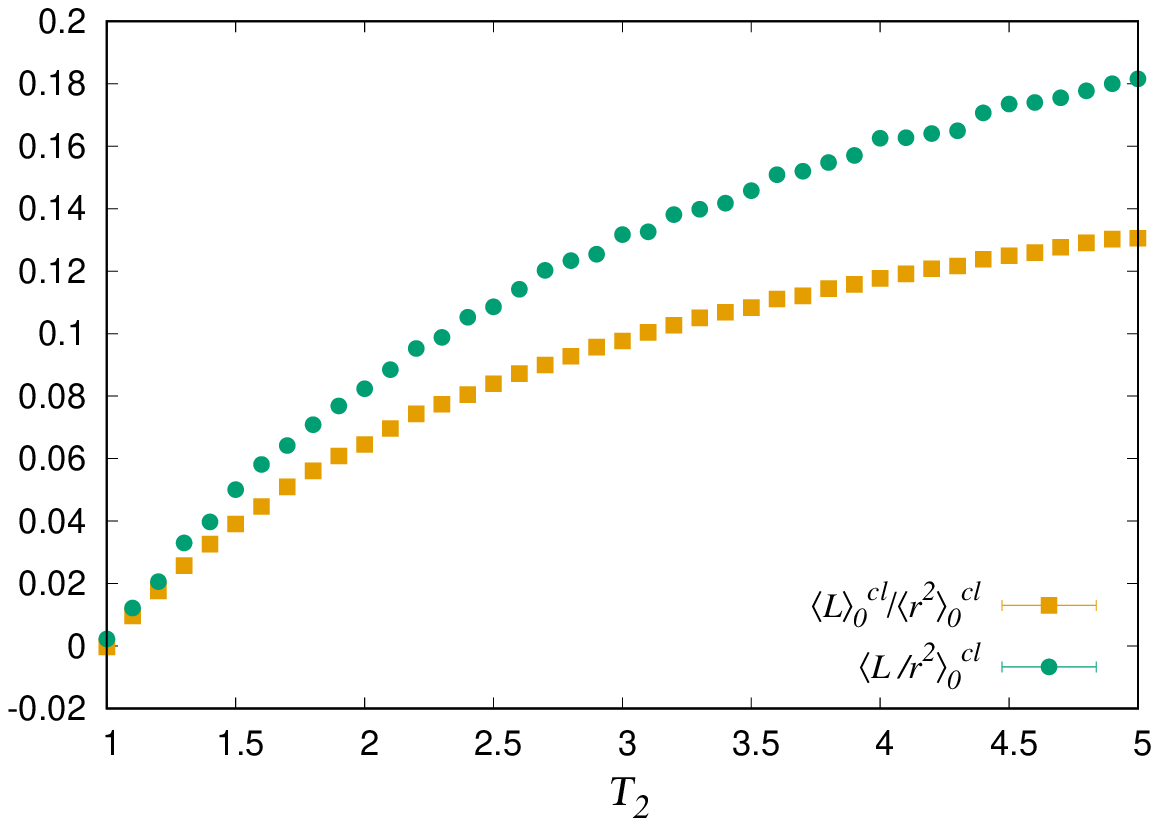}
\caption{Left Panel: $\average{L/r^2}_0$ and $\average{L}_0/ \average{r^2}_0$ as  functions of the potential rotation angle $\alpha$, as obtained by integrating the classical Langevin equation with $10^5$ independent trajectories of time duration $t_f=1$, $T_1=1$, $T_2=4$, $u_1=1,\, u_2=1/4, \, m=\eta=1$. Right Panel:  $\average{L/ r^2}_0$ and $\average{L}_0/ \average{r^2}_0$ as  functions of the temperature $T_2$ with $t_f=10$, all the other parameters as in the left panel.}
\label{fig:check}
\end{figure}
%
%%%%%%%%%%%%%%
\subsection{Time dependent force}

In the presence of an external periodic drive the quantum rotator is locked onto a limit cycle.
Choosing a drive with amplitude $D$ and frequency $\omega_0$,
%
%\begin{eqnarray}
%&&f_1(t)=D\cos\omega_0t,
%\label{af1}
%\\
%&&f_2(t)=D\sin\omega_0t,
%\label{af2}
%\end{eqnarray}
%
\begin{equation}
\boldsymbol f(t)=D(\cos\omega_0t,\sin\omega_0t),
\label{af}
\end{equation} 
we have
\begin{eqnarray}
&&\langle L\rangle=\langle L\rangle_0+\langle L\rangle_D,
\\
&&\langle L\rangle_D= D^2 \omega_0 K(\omega_0)=
mD^2\omega_0\frac{(A-m\omega_0^2)(B-m\omega_0^2)+(\omega_0\eta)^2-C^2}{|Z(\omega_0)|^2}.
\end{eqnarray}
The fluctuating moment of inertia is given by $I=m(x_1^2+x_2^2)$.
For the mean moment of inertia we have, setting $\langle I\rangle=\langle I\rangle_0+\langle I\rangle_D$,
\begin{eqnarray}
\langle I\rangle_0=&&m\eta \int\frac{d\omega}{2\pi}\hbar\omega
\bigg[\frac{(B-m\omega^2)^2+(\omega\eta)^2+C^2}{|Z(\omega)|^2}\coth(\hbar\omega/2T_1)
\nonumber
\\
&&~~~~~~~~~~~~~~+\frac{(A-m\omega^2)^2+(\omega\eta)^2+C^2}{|Z(\omega)|^2}\coth(\hbar\omega/2T_2)
 \bigg],
\\
\langle I\rangle_D=&&\frac{mD^2}{2}\frac{(A-m\omega_0^2)^2+(B-m\omega_0^2)^2+2(\omega_0\eta)^2+2C^2}
{|Z(\omega_0)|^2}.
\end{eqnarray}
For the ratio $\langle L\rangle_D/\langle I\rangle_D$ we obtain
\begin{eqnarray}
\frac{\langle L\rangle_D}{\langle I\rangle_D}=
2\omega_0\frac{(A-m\omega_0^2)(B-m\omega_0^2)+(\omega_0\eta)^2-C^2}
{(A-m\omega_0^2)^2+(B-m\omega_0^2)^2+2(\omega_0\eta)^2+2C^2}.
\end{eqnarray}
In the limit of $\omega_0$ large the system is stiff and we obtain $\langle L\rangle_D=
\langle I\rangle_D\omega_0$, characteristic of a rigid body \cite{Landau59b}.
%%%%%%%%%%%%%%%%%%%%%%%%%
\section{\label{app3}Work and heat rates}
%%%%%%%%%%%%%%%%%%%%%%%%%
From the equations of motion (\ref{aqlan}) the fluctuating forces associated with the heat reservoirs are given
by $(-\eta\dot x_n(t)+\xi_n(t))$. For the rates of heat acting on the rotator we then have
$r_{q_n}=(\dot x_n(-\eta\dot x_n(t)+\xi_n(t)))_W$. Inserting eq.~(\ref{aqlan}) we obtain for the fluctuating heat rates
\begin{eqnarray}
&&r_{q_1}=(\dot x_1(m\ddot x_1 +A x_1+C x_2-f_1))_W,
\label{aheat1}
\\
&&r_{q_2}=(\dot x_2(m\ddot x_2 +B x_2+C x_1-f_2))_W.
\label{aheat2}
\end{eqnarray}

The Hamiltonian in eq.~(\ref{aham1}) depends explicitly on time through $(f_1(t),f_2(t))$.
As a result $\dot H=\sum_n((\partial H_0/\partial p_n)\dot p_n+\sum_n((\partial H_0/\partial x_n)\dot x_n-\sum_n x_n\dot f_n$, where the first two terms correspond to the total fluctuating heat rate $r_{q_1}+r_{q_2}$, 
while the third term is the fluctuating rate of work done on the system
\begin{eqnarray}
r_w=-x_1\dot f_1-x_2\dot f_2.
\label{awork}
\end{eqnarray}
In a steady averaged state we have $\langle\dot H\rangle=0$ and we infer the conservation law
$\langle r_{q_1}+r_{q_2}\rangle + \langle r_w\rangle =0$. This result can be obtained by direct calculation, as detailed below. 

Subject to the periodic drive in eq.~(\ref{af}) 
we obtain by insertion of eqs.~(\ref{asol1}) and (\ref{asol2}) and averaging  the mean rate of work
\begin{eqnarray}
\langle r_w\rangle=\frac{\eta}{2}D^2\omega_0^2\frac{(A-m\omega_0^2)^2+(B-m\omega_0^2)^2+2(\omega_0\eta)^2+2C^2}
{|Z(\omega_0)|^2}.
\label{awork2}
\end{eqnarray}
The non equilibrium heat transfer rate is given by $\langle\Delta r_q\rangle=\langle r_{q_1}\rangle-\langle r_{q_2}\rangle$.
From the definition $\langle L\rangle=\langle m(x_1\dot x_2-x_2\dot x_1)\rangle_W$ and the definitions in 
eqs.~(\ref{aheat1})
and (\ref{aheat2}) we obtain the general identity
\begin{eqnarray}
\langle\Delta r_q\rangle=-\frac{C}{m}\langle L\rangle-\langle\dot x_1\rangle f_1+\langle\dot x_2\rangle f_2.
\label{adelheat}
\end{eqnarray}
For the periodic drive in particular we find
\begin{eqnarray}
\langle\Delta r_q\rangle=-\frac{C}{m}\langle L\rangle_0
+\frac{\eta}{2}(D\omega_0)^2\frac{(A-m\omega_0^2)^2-(B-m\omega_0^2)^2-4C\eta\omega_0}
{|Z(\omega_0)|^2}.
\label{adelheat2}
\end{eqnarray}
For the individual heat transfer rates $r_{q_1}$ and $r_{q_2}$ we have
\begin{eqnarray}
&&\langle r_{q_1}\rangle=-\frac{C}{2m}\langle L\rangle_0-\frac{\eta}{2}(D\omega_0)^2
\frac{(B-m\omega_0^2)^2+(C+\omega_0\eta)^2}{|Z(\omega_0)|^2},
\\
&&\langle r_{q_2}\rangle=+\frac{C}{2m}\langle L\rangle_0-\frac{\eta}{2}(D\omega_0)^2
\frac{(A-m\omega_0^2)^2+(C-\omega_0\eta)^2}{|Z(\omega_0)|^2}.
\end{eqnarray}

\end{widetext}

\end{document}